\documentstyle[aps,eqsecnum,preprint,tighten]{revtex}
\input epsf
\newcommand\LQCD{{\Lambda_{\rm QCD}}}
\def\INSERTFIG#1#2#3{
\multiply\baselineskip
by 3 \divide\baselineskip by 4
\vbox{\bigskip \vbox{\hfil\epsfbox{#1}\hfill}%
{
\noindent%
{FIG.\ #2. }{\sl #3 \medskip}}
}\multiply\baselineskip by 4 \divide\baselineskip by 3 \hfill}%
\begin{document}
\draft
\preprint{\vbox{\hbox{UCSD/PTH 99--04}}}

\title{Determining $V_{ub}$ from $B^+ \to D^{*+}_s e^+e^-$ 
and $B^+ \to D^{*+} e^+e^-$ }

\author{David H. Evans\footnote{daevans@physics.ucsd.edu}, Benjam\'{\i}n
Grinstein\footnote{bgrinstein@ucsd.edu}
 and Detlef R. Nolte\footnote{dnolte@ucsd.edu}}

\address{Department of Physics,
University of California at San Diego, La Jolla, CA 92093}

\date{March, 1999}
  
\maketitle

\widetext

\begin{abstract}
It was recently pointed out that the decays $B^+ \to D^{*+}_s \gamma$
and $B^+ \to D^{*+} \gamma$ can be used for an extraction of
$|V_{ub}|$.  The theory of these decays is poorly understood. It was
shown that in a world of almost degenerate $b$ and $c$-quarks
the decay would be computable. The severe difficulties that are
encountered in the realistic calculation stem primarily from the very
hard photon produced in the two body decay. We point out that in the
decays $B^+ \to D^{*+}_s e^+e^-$ 
and $B^+ \to D^{*+} e^+e^-$ the photon vertex is soft when the charmed
meson is nearly at rest (in the $B^+$ rest frame). This allows us to
compute with some confidence the decay rate in a restricted but
interesting kinematic regime. Given enough data the extraction of
$V_{ub}$ with reasonably small uncertainties could proceed through an
analysis of these exclusive decays much as is done in the
determination of $V_{cb}$.

\end{abstract}


\narrowtext

\section{Introduction}

The determination of the Kobayashi-Maskawa (KM) matrix element $V_{ub}$ is
one of the central themes of particle physics today. The verification
of the KM theory of CP violation in the six quark model requires an
independent measurement of the magnitude and phases of the KM matrix
elements. The physics community has invested heavily  in
$B$-factories, facilities devoted to the measurement of these
magnitudes and phases. However, the extraction of $V_{ub}$ from data
requires calculation of  hadronic transition rates that are marred by
the usual host of difficulties introduced by the strong interactions.

For example, the first determination of $V_{ub}$ has been made in the
inclusive semileptonic decay of $B$-mesons. The inclusive decay rate,
appropriately smeared over kinematic variables, can be reliably
estimated\cite{chay}. Since $|V_{cb}|^2\gg|V_{ub}|^2$, the rate is dominated by
the decay into charmed mesons. To isolate the contribution from
$V_{ub}$ measurements are made of the decay rate for electron or muon energies
that are large enough to exclude contamination from the decay into
charm. However, in this very restricted kinematic range the
theoretical prediction is highly unreliable, and it is known to be
dominated by incalculable non-perturbative effects\cite{incl}.

A novel procedure offers more hope. If one can reconstruct the
neutrino in the semileptonic decay of $B$ mesons only small
invariant hadronic masses are required to suppress or eliminate the
contribution from decays into charmed particles\cite{hadrmass}. The
invariant mass
spectra are under better theoretical control because the allowed final
states are more numerous, so quark-hadron duality is more likely to be
valid than in the restricted region of very high electron or muon
energies.

In this paper we investigate the extraction of $V_{ub}$ from an
exclusive decay mode of charged $B$-mesons. Exclusive semileptonic
decay modes, such as $B\to\pi\ell\bar\nu$ or $B\to\rho\ell\bar\nu$,
have proved difficult to study theoretically. The rates are described
by non-perturbative form factors which are difficult to estimate.  The
exclusive decay mode $B^+ \to D^{*+} e^+e^-$ depends on
non-perturbative dynamics too. But as we will see, under reasonable
theoretical assumptions, the non-perturbative dynamics can be
extracted from other measurements.

A similar idea was recently proposed in the description of the decays
$B^+ \to D^{*+}_s \gamma$ and $B^+ \to
D^{*+}\gamma$\cite{leb99}. There the transition was assumed to be
dominated by the lowest poles ($B^*$ and $D_s$ or $D$) and was given
in terms of the amplitude for $B\bar B$ mixing and the probability
rate for $D^*\to D\gamma$. The validity of the symmetry arguments used
there relied heavily on a theoretical limit, that $m_b-m_c\alt \LQCD$,
which is not likely to be a good approximation to reality. This
assumption was important in relating the hard photon emission in $B^+
\to D^{*+} \gamma$ to the soft emission in $D^* \to D \gamma$. We
observe in this paper that for the decay into a lepton pair, rather
than a photon, the extended phase space includes a region where the
$D^*$ meson is nearly at rest in the $B$ rest-frame, in which the
photon is soft. Consequently the unphysical assumption $m_b-m_c\alt
\LQCD$ can be dropped.

We make several assumptions and approximations in our
calculation of the rates for $B^+ \to D^{*+}_s e^+e^-$ 
and $B^+ \to D^{*+} e^+e^-$. If the assumptions prove to hold and the
approximations are reliable a determination of $V_{ub}$ through
these exclusive modes could proceed in much the same fashion as that
of $V_{cb}$ through the exclusive semileptonic mode $B\to
D^*\ell\bar\nu$.  Theory reliably predicts the rate in the
kinematic region in which the $D^*$ is produced at rest in the $B$
rest-frame. Experimentally the rate is measured away from this point
and then extrapolated. The procedure would be entirely analogous for
the much rarer, but cleaner, decay we study,  $B^+ \to D^{*+}
e^+e^-$.

\section{Basic Assumptions And Symmetries}
Before we present our calculation we discuss in
detail the nature of the assumptions and approximations that we use.
To lowest order in
electromagnetic and weak interactions, the amplitude for the process
$B^+ \to D^{*+} \gamma^*$ is given in terms of
\begin{equation}
\langle D^{*+}| \int d^4x\;e^{iq\cdot x}
\; T(j^\mu_{\rm em}(x){\cal O}(0)) |B^+\rangle.
\end{equation}
Here ${\cal O}=\bar b\gamma^\nu(1-\gamma_5) u
\bar c\gamma^\nu(1-\gamma_5) d$ is a four quark operator responsible
for the weak transition, $j^\mu_{\rm em}$ is the electromagnetic
current and $T$ stands for time ordering of these operators. For each
time ordering one may insert a complete set of states. Our first assumption
is that the lowest mass states dominate, giving
\begin{eqnarray}
\label{eq:insertstates}
\int\!\! d\Phi& &\!\!\!\!\!\!\!\!
 \int\!\! d^4x\;
e^{iq\cdot x} (\theta(x^0)\langle D^{*+}| j^\mu_{\rm em}(x)
|D^+\rangle\langle D^+|
{\cal O}(0)|B^+\rangle\nonumber \\
&+&
\theta(-x^0)\langle D^{*+}| {\cal O}(0)
|B^{*+}\rangle\langle B^{*+}|
j^\mu_{\rm em}(x) 
|B^+\rangle),
\end{eqnarray}
where $\int d\Phi$ is the one body phase space. We know of no reliable
way to test this assumption. Much of heavy quark chiral perturbation
theory\cite{Wise} relies on analogous assumptions. The effect of
excited heavy mesons in loops in chiral perturbation theory has been
analyzed and the result seems to be sensitive to the precise
values of masses and couplings\cite{falk93}. This is an area
where more research is needed before firm conclusions can be drawn.

Given this assumption, the transition amplitude is calculated in terms
of the matrix elements for the weak operator
\begin{equation}
\label{O-matrixelements}
\langle D^+| {\cal O}(0)|B^+\rangle 
\qquad\hbox{and}\qquad
\langle D^{*+}| {\cal O}(0) |B^{*+}\rangle
\end{equation}
and the matrix elements of the electromagnetic current,
\begin{equation}
\label{em-matrixelements}
\langle D^{*+}| j^\mu_{\rm em}(0) |D^+\rangle
\qquad\hbox{and}\qquad
\langle B^{*+}| j^\mu_{\rm em}(0) |B^+\rangle.
\end{equation}
The matrix element of a vector current between a pseudoscalar and a
vector meson can be written in terms of a single real form factor,
which can be defined as follows
\begin{equation}
\label{eq:formfdefd}
\langle {\bf p', \epsilon}| j^\mu_{\rm em}(0) |{\bf p}\rangle =
ig^{\mbox{ }}_{X}(t)\epsilon^{\mu\nu\lambda\sigma}p_\nu
p^\prime_\lambda\epsilon^*_\sigma.
\end{equation}
Here $X=B$ or $D$ identifies the one particle states, $\epsilon$ is
the polarization of the vector meson, and the time component of the
four vectors $p$ and $p'$ is the energy of the state,
$p_0=\sqrt{M^2+|{\bf p}|^2}$. The Lorentz invariant form factor
$g^{\mbox{ }}_X$ is a function only of the invariant momentum transfer
$t=(p-p')^2$. Heavy quark flavor symmetry\cite{HQET} can be used to
relate $g^{\vphantom{\dagger}}_B$ and $g^{\vphantom{\dagger}}_D$ when
the argument is $t\approx0$. Moreover, $g^{\mbox{ }}_X(0)=\mu^{\mbox{
}}_X$, where $\mu^{\mbox{ }}_X$ is the magnetic moment defined in
Ref.~\cite{amundson} determines the radiative decay rate of the $D^*$
meson\cite{amundson},
\begin{equation}
\label{eq:dstartodgamma}
\Gamma(D^*\to D\gamma)=
\frac\alpha3|\mu^{\mbox{ }}_X|^2
\left(\frac{M_{D^*}^2-M_D^2}{2M_{D^*}}\right)^3.
\end{equation}
The heavy quark contribution to the magnetic moment of the meson is
suppressed in the large mass limit, so $\mu^{\mbox{ }}_X=\mu Q_q$,
where $Q_q$ is the charge of the light quark or anti-quark in the
heavy meson. Therefore $\mu^{\mbox{ }}_{D^+}\approx\frac12\mu^{\mbox{
}}_{B^+}$. 

The functional form of the form factor $g(t)$ is not known. Clearly
the dimensionful scale that determines its behavior is on the order of
the mass of the lightest vector meson, the $\rho$-meson. Now, in the
rest-frame of the decaying $B^+$ meson, ${\bf p}=0$, and
$t=(p'_0-M)^2-|{\bf p'}|^2=(M'-M)^2-2M(p'_0-M')$. So at ${\bf p'}=0$
the momentum transfer variable is small and positive,
$t\approx(50~\hbox{MeV})^2$ for $B$ mesons and
$t\approx(150~\hbox{MeV})^2$ for $D$ mesons. As $|{\bf p'}|$
increases $t$ goes through zero and turns negative. Therefore
there is a region of phase space, close to the point of zero recoil of
the $D^*$ meson, for which the matrix elements in
(\ref{em-matrixelements}) are known directly from experiment or via
heavy quark symmetry.

It is not obvious how the form factors $g(t)$ should be incorporated
in the calculation. Using the matrix element given in terms of a form
factor in Eq.~(\ref{eq:formfdefd}) in the expression for the amplitude
in Eq.~(\ref{eq:insertstates}) gives non-covariant results. For a
constant or linear form factor a covariant result is recovered if the
``Z'' graphs are included (ie, for $x^0<0$ a graph with an
intermediate state of the three particles $D^-$, $D^{*+}$ and
$B^+$). But covariance is lost for more general form factors. Although
the ``Z'' graphs are needed to recover covariance, the behavior of the
form factor suggests that they are negligible in practice. In what
follows we will compute in the rest frame of the decaying $B^+$ meson
and neglect ``Z'' graphs. This is a good approximation at least when
the resulting $D^*$ meson is at rest or recoiling slowly. The question
of how to  incorporate the form factor effects consistently is an open
one and we hope to return to it in the future.

Turning our attention to the weak matrix elements in
(\ref{O-matrixelements}) we notice immediately the close resemblance
of these to the matrix elements in $B\bar B$ mixing,
\begin{equation}
\label{eq:BBdefd}
\langle \bar B|{\cal O}_B| B\rangle=\frac83f_B^2M_B^2B^{\vphantom2}_B.
\end{equation}
Here ${\cal O}_B =\bar b\gamma^\nu(1-\gamma_5) d
\bar b \gamma_\nu(1-\gamma_5) d$ and the right hand side parameterizes
the matrix element in terms of $B_B$ and the mass $M_B$ and decay
constant $f_B$ of the $B$-meson. In both
cases a heavy quark and a light quark are annihilated at a local
vertex where a heavy anti-quark and a light quark are created. In
$B\bar B$ mixing the heavy anti-quark  is produced  at rest. In
$B^+\to D^* \ell^-\ell^+$ we already singled out the case of zero
recoil as the one where we can compute the electromagnetic
transition. The crudest approximation, vacuum insertion, works
surprisingly well in the case of $B$-mesons, at least as indicated by
calculations in quenched lattice QCD. Vacuum insertion gives
\begin{equation}
\label{eq:vacinsert}
\langle D^+| {\cal O}(0)|B^+\rangle =f^{\vphantom\dagger}_B
f^{\vphantom\dagger}_D M^{\vphantom\dagger}_B
M^{\vphantom\dagger}_D,
\end{equation}
and the analogous result for the vector mesons. 

One may try a more sophisticated attempt at predicting this matrix
element, using symmetries to relate it to the measured matrix element
for $B\bar B$ mixing. While heavy quark symmetries do allow us to
exchange the heavy $b$ anti-quark for a heavy $c$ anti-quark, isospin
(or $SU(3)$, for the $D^*_s$) which is needed to exchange the light
quarks does not give a relation between the matrix elements. The
operator ${\cal O}$ transforms in the reducible ${\bf 3\oplus 1}$
representation of isospin (${\bf 6\oplus \bar 3}$ representation of
$SU(3)$), but the corresponding operator for $B\bar B$ mixing
transforms in the irreducible ${\bf 3}$ of isospin (${\bf 6}$ of
$SU(3)$). We make the plausible assumption that the matrix element of
the color octet operator ${\cal O}_8=\bar bT^a\gamma^\nu(1-\gamma_5) u
\bar cT^a\gamma_\nu(1-\gamma_5) d$ is negligible. Indeed, if one could
insert a complete set of (gauge invariant) states between the currents
in ${\cal O}_8$, gauge invariance would immediately imply that the
matrix element vanishes. Inserting a complete set of states seems less
drastic than assuming saturation by the vacuum alone. The assumption
that ${\cal O}_8$ is negligible leads to relations between the matrix
elements of the ${\bf 3}$ and ${\bf1}$ components of ${\cal O}$. With
$P_\pm\equiv1\pm\gamma_5$, define
\begin{equation}
{\cal O}_\pm=\bar b\gamma^\nu P_- u \bar
c\gamma^\nu P_- d \pm
\bar b\gamma^\nu P_- d \bar
c\gamma^\nu P_- u
\end{equation}
then ${\cal O}_+$  and ${\cal O}_-$ transform in the {\bf3} and {\bf1}
representations of isospin and we have the change of basis relations:
\begin{equation}
\pmatrix{{\cal O} \cr {\cal O}_8}=
\pmatrix{\frac12&\frac12\cr \frac16&-\frac13}
\pmatrix{{\cal O}_+ \cr {\cal O}_-}
\end{equation}
The vanishing of the matrix elements of ${\cal O}_8$ implies that
between $B^+$ and $D^{*+}$ states  $\langle
{\cal O}_-\rangle=\frac12 \langle {\cal O}_+\rangle$ and therefore 
$\langle {\cal O}\rangle=\frac34\langle{\cal O}_+\rangle$.

In order to relate the matrix elements of ${\cal O}_B$ and ${\cal O}$
we also need to make use of heavy quark symmetries. The first step is
to write operators in a heavy quark effective theory (HQET) $\tilde
{\cal O}_B$ and $\tilde {\cal O}$ corresponding, respectively, to the
operators ${\cal O}_B$ and ${\cal O}$ in the original theory. If
$h_{\bar Q}$ is an HQET field that destroys a heavy quark of
four-velocity $v$, and $h_Q^\dagger$ an HQET field that creates a
heavy antiquark with the  same velocity $v$ then
\begin{equation}
\label{eq:tildeOBdefd}
{\cal O}_B \rightarrow \tilde{\cal O}_B =
  2\;h_{\bar b}^T\gamma^0\gamma^\nu P_-  d\,
\bar h_b^{\vphantom{t}} \gamma_\nu P_-  d.
\end{equation}
The superscript $T$ stands for the transpose. The factor of two arises
from the two ways of obtaining a heavy quark creation and
corresponding anti-quark annihilation operators. Similarly
\begin{eqnarray}
{\cal O}(x) &\rightarrow& \tilde{\cal O}(x) =e^{-i(m_b-m_c)v\cdot
x}\nonumber\\
&\times&  h_{\bar b}^T(x)\gamma^0\gamma^\nu P_-  u(x)\,
\bar h_c(x) \gamma_\nu P_-  d(x),
\end{eqnarray}
and analogous correspondences for ${\cal O}_8$ and  ${\cal O}_\pm$.
In the HQET there is a $SU(2)\times SU(2)$ flavor symmetry that
rotates $b\leftrightarrow c$ and $\bar b\leftrightarrow \bar c$
independently. The symmetry is explicit between states with mass
independent normalization. Using this together with suppression of
${\cal O}_8$ and isospin symmetry we have
\begin{equation}
\frac{\langle D^+|\tilde{\cal O}|B^+\rangle
}{\sqrt{M^{\vphantom\dagger}_BM^{\vphantom\dagger}_D}}
=\frac34\;\frac{\langle D^+|\tilde{\cal O}_+|B^+\rangle
}{\sqrt{M^{\vphantom\dagger}_BM^{\vphantom\dagger}_D}} 
=\frac12\;
\frac34\;\frac{\langle \bar B^0|\tilde{\cal O}_B|B^0\rangle
}{{M^{\vphantom\dagger}_B}}.
\end{equation}
The factor of $1/2$ in the last step is a Clebsch-Gordan coefficient.
Note that the factor of $2$ appearing in Eq.~(\ref{eq:tildeOBdefd})
was retained in the definition of $\tilde {\cal O}_B$ so that $\tilde
{\cal O}_B$ and $\tilde {\cal O}_+$ have the same normalization in the
triplet tensor operator.  In terms of $B_B$ defined in
Eq.~(\ref{eq:BBdefd}) we finally obtain for states with vanishing
momentum 
\begin{equation}
\langle D^+|{\cal O}(x)|B^+\rangle = e^{-i(m_b-m_c)t}
\; \sqrt{\frac{M_D}{M_B}}f_B^2M_B^2B^{\vphantom2}_B,
\end{equation}
or, using $f_B^2M_B=f_D^2M_D$  from heavy quark symmetry, and
evaluating at the origin
\begin{equation}
\label{eq:mixing-result}
\langle D^+|{\cal O}(0)|B^+\rangle = f^{\vphantom{\dagger}}_B 
f^{\vphantom{\dagger}}_D M^{\vphantom{\dagger}}_B
M^{\vphantom{\dagger}}_D B^{\vphantom{\dagger}}_B.
\end{equation}
Comparing with Eq.~(\ref{eq:vacinsert}) we see that the symmetry
argument and the octet suppression assumption give the result that the
$B$ correction factor for $B^+\to D^+$ is precisely
$B^{\vphantom{\dagger}}_B$, the correction factor for $B\bar B$
mixing.

The mixing between $B^{*+}$ and $D^{*+}$ in (\ref{O-matrixelements})
is fixed by heavy quark spin symmetry\cite{GJMSW} to have the same
magnitude but opposite sign than Eq.~(\ref{eq:mixing-result}). As a
consequence of this sign the contributions to the rate from the two
time orderings in Eq.~(\ref{eq:insertstates}) add. However, in the
decay to a pseudoscalar, $B^+\rightarrow D^+e^+e^-$, the two time
orderings tend to cancel each other. Moreover, the electromagnetic
form factors in $B^+\rightarrow D^+e^+e^-$ are unity at $t=0$, so the
cancellation is rather good. For this reason we have not analyzed this
process further.

\section{Results and Discussion}
The effective Hamiltonian for the weak transition is modified by short
distance QCD corrections,
\begin{equation}
{\cal H}'_{\rm eff}= \frac{G_F}{\sqrt2}\,V_{ub}V^*_{cq}\left(
c(\mu/M_W){\cal O}+c_8(\mu/M_W){\cal O}_8\right).
\end{equation}
The short distance coefficients $c$ and $c_8$ are chosen so that their
dependence on the renormalization point $\mu$ cancels the
$\mu$-dependence of operators, in such a way that matrix elements of
the effective Hamiltonian are $\mu$-independent. Resuming the leading
logs, they are given at $\mu=m_b$ in terms of
$x=\left(\alpha_s(m_b)/\alpha_s(M_W)\right)^{6/23}$ by
$c=\frac13x^2+\frac23x^{-1}$ and $c_8=x^{-1}-x^2$; numerically, with
$\alpha_s(m_b)/\alpha_s(M_W)\approx1.9$ one has $c\approx1.0$ and
$c_8\approx-0.6$.

The calculation is now straightforward. The transition amplitude for
$B^+\to D^{*+}e^+e^-$ is given by the vector in
Eq.~(\ref{eq:insertstates}) contracted with the leptonic current and
multiplied by the photon propagator and coupling constants. The
hadronic matrix elements in Eq.~(\ref{eq:insertstates}) are computed
as described in the previous section. 

There is one subtlety.  For the first time ordering
in Eq.~(\ref{eq:insertstates}) the energy denominator is
$q_0+E_{D^*+}-E_D=M_B-E_D$, where $E_D=\sqrt{E_B^2-M_B^2+M_D^2}$ is
the energy of the intermediate $D$-meson. For the second time ordering
the energy denominator is $q_0-E_B+E_{B^*}=-E_{D^*}+E_{B^*}$, where
the intermediate $B^*$ energy is
$E_{B^*}=\sqrt{E_{D^*}^2-M_{D^*}^2+M_{B^*}^2}$. In the rest frame of
the $B$-meson and at zero $D^*$ recoil, these denominators are
$M_B-M_D$ and $-(M_B-M_D) +O(1/M)$. For non-zero $D^*$ recoil, the
denominators are different. Therefore our results do not  depend on
the form factors through the simple combination  $g_B(t_B)+g_D(t_D)$,
except at the point of zero recoil.

\def\SKIP#1{}
\SKIP{
\epsfxsize=6in
\INSERTFIG{Fofq.eps}{1}{Plot of the dimensionless function ${\cal F}(q^2)$ that
appears in the rate for $B^+\rightarrow D^{*+}e^+e^-$ decay,
Eq.~(\ref{eq:decayrate}), with $q^2$ in GeV.} 
}

The differential decay rate is best written in terms of two functions
that contain the kinematic dependence. The dependence on the vector current
form factors is in
\begin{equation}
{\cal G}(q^2)=N_{\cal G}^{-1}
\left(g_D(t_D)+\frac{M_B(M_B-M_D)}{E_{B^*}(E_{B^*}-E_{D^*})}g_B(t_B)\right),
\end{equation}
where $t_X=(M_X-M_{X^*})^2-2M_X(E_{X^*}-M_{X^*})$ and $N_{\cal G}$ is a
normalization factor defined so that ${\cal G}=1$ at $q^2=q^2_{\rm
max}=(M_B-M_{D^*})^2$ and is approximately
\begin{equation}
N_{\cal G}\approx\mu_D+\mu_B\approx3\mu_D.
\end{equation}
The second dimensionless function is a combination of amplitude and phase
space dependence.
It vanishes at the upper end-point as $(q^2_{\rm max}-q^2)^{3/2}$. It
is given by
\begin{eqnarray}
{\cal F}(q^2)& = &\frac{
\sqrt{1-4m_e^2/q^2}(1+2m_e^2/q^2)}{q^2M_B^4}\\
&\times&(q^2 - (M^{\vphantom{W}}_B+M{\vphantom{W}}_D)^2)^{3/2}
( q^2 - (M^{\vphantom{W}}_B-M{\vphantom{W}}_D)^2)^{3/2}.\nonumber
\end{eqnarray}
In terms of these functions we finally obtain 
\begin{equation}
\label{eq:decayrate}
\frac{d\Gamma}{dq^2} = \frac{3\alpha^2}{64\pi}G_F^2|V_{ub}V_{cq}|^2c^2
\frac{f_B^4M_B^2M_DB_B^2}{(M^{\vphantom{W}}_B-M{\vphantom{W}}_D)^2}
\mu^2_D{\cal G}^2(q^2){\cal F}(q^2).
\end{equation}
Using Eq.~(\ref{eq:dstartodgamma}) 
the differential branching fraction for $B^+\rightarrow
D^{*+}e^+e^-$ is, numerically,
\begin{eqnarray}
\tau_BM_B^2\frac{d\Gamma}{dq^2}&=&1.3\times10^{-11}
\left(\frac{|V_{ub}V_{cd}|}{8\times10^{-4}}\right)^2
\left(\frac{f_B\sqrt{B_B}}{170~\hbox{MeV}}\right)^4\nonumber\\
&\times&\left(\frac{\Gamma(D^*\rightarrow D\gamma)}{4.2~\hbox{KeV}}\right)
{\cal G}^2(q^2){\cal F}(q^2),
\end{eqnarray}
and about a factor of 16 bigger for $B^+\rightarrow
D^{*+}_se^+e^-$. The integrated branching fraction, restricted to
$q^2>0.1~{\rm GeV}$, is $6.3\times10^{-10}$ assuming ${\cal G}$ is
constant, but for ${\cal G}$ given by a single $\rho$-pole, 
${\cal G}=m_\rho^2/(m_\rho^2+q^2-q^2_{\rm max})$, it is
reduced to $2.2\times10^{-12}$ (and, again, a factor of 16 bigger for
$B^+\to D^{*+}_se^+e^-$).

\section{Conclusions}
We have calculated the rates for  $B^+\rightarrow D^{*+}e^+e^-$ and 
$B^+\rightarrow D^{*+}_se^+e^-$. Our primary motivation is the 
observation that at the kinematic point where the $D^*_{(s)}$ meson is
recoiling slowly in the $B$ rest frame the amplitude can be computed
in terms of other observables, like the width of the $D^*$ and the
$B-\bar B$ mixing rate. While this observation opens up exciting possibilities
of computation and of the determination of $V_{ub}$, we leave many
loose ends. The computation makes several assumptions
which need to be further tested. Among these the assumption that the
rate is dominated by the lowest mass resonance is most suspect. 
Also, we leave unanswered a fundamental question: how can one
covariantly incorporate the electromagnetic form factors
$g(t)$, which are here assumed to be known,  into the  calculation?

We are not discouraged by the seemingly small branching fraction
obtained. The theoretical questions and challenges raised (see
previous paragraph) are worth pursuing in their own right. Their
resolution may guide us to a better method for a determination of
$V_{ub}$ in exclusive decays.
 
{\it Acknowledgments} This work is supported by the Department of
Energy under contract No.\ DOE-FG03-97ER40546.

\end{document}